# AC-Driven Perovskite Light-Emitting Transistors[†]

**Francesco Maddalena,[1] Xin Yu Chin,[2] Daniele Cortecchia,[2,3] Annalisa Bruno,[2] Cesare Soci[1,4,*]**

[1.] *Division of Physics and Applied Physics, School of Physical and Mathematical Sciences, Nanyang Technological University, Singapore 637371*

[2.] *Energy Research Institute @ NTU, Nanyang Technological University, Singapore 637553*

[3] *Interdisciplinary Graduate School, Nanyang Technological University, Singapore 639798*

[4] *Centre for Disruptive Photonic Technologies, TPI, Nanyang Technological University, Singapore 637371*

[†] *Electronic supplementary information (ESI) available.*

**Perovskite light-emitting field-effect transistors (PeLEFETs) provide a versatile device architecture to control transport and electroluminescence properties of hybrid perovskites, enabling injection of high charge carrier density and spatial control of the radiative recombination zone. Ionic screening and organic cation polarization effects typical of metal-halide perovskites, however, critically affect PeLEFET efficiency and reliability. In this work, we demonstrate a new device operation mode based on high-frequency modulation of the applied voltages, which allows significant reduction of ionic drift/screening in methylammonium lead iodide light-emitting transistors. In optimized top contact PeLEFETs, AC operation results in brighter and more uniform electroluminescence compared to DC-driven devices. Moreover, high frequency modulation enables electroluminescence emission up to room temperature.**

**Keywords:** Metal-halide perovskites, methylammonium lead iodide, light-emitting field-effect transistors, AC-driven light-emitting devices.

# INTRODUCTION

Solution-processable hybrid organic–inorganic halide perovskites (HOIPs) are a class of materials with remarkable features such as high photoluminescence efficiencies,[1-2] long carrier lifetimes and diffusion lengths,[3-5] higher mobility than typical solution processed materials,[5-10] low thresholds in optical pumped lasing,[1, 11] tunability of the optical band-gap from the visible region to the near infrared[6, 12-13] and white light emission.[14-15] HOIPs have shown extremely good performance in photovoltaic devices,[16-18] photo-detectors,[19] X-ray scintillation detectors,[20] and light-emitting diodes.[2, 21-24] HOIPs are also promising semiconductors for field-effect transistors,[9, 25-30] and recently we demonstrated perovskite light-emitting field-effect transistors (PeLEFETs), operating at low temperatures.[9] LEFETs offer several advantages compared to LEDs, such as very high charge carrier densities and controllable current flow, charge injection and emission patterns, leading to electroluminescence efficiencies that can outperform equivalent LEDs.[31] Moreover, electroluminescence and electrical switching are integrated into the device architecture itself.

Despite their great advantages, hybrid perovskites are severely affected by ionic motion[9, 32-34] and polarization disorder of the organic cations,[35-39] which reduce operational stability, reliability and reproducibility of perovskite-based devices. These phenomena give rise to undesirable side-effects, such as electrical hysteresis[40-42] during applied bias conditions[43-46] and low *effective* charge carrier mobility in devices, orders of magnitude lower than the theoretical values or hall-effect measurements.[5, 9, 25, 47-49] In particular, these effects are responsible for poling-dependent efficiency of perovskite solar cells,[43] unreliable operation of perovskites LEDs, especially at higher operational voltages,[50] and complete loss of gating effects at room temperature in perovskite field-effect transistors.[9] Ionic transport is mainly due to the presence of ion vacancies in the perovskite, which allow ions to move within the crystalline structure of the material.[32] The most probable source of ionic drift in methylammonium lead iodide is the iodine anion.[32] but it has been shown that cations can also contribute.[51] The disorder introduced by organic cations is caused by cation dipole fluctuations,[30, 39, 52-54] which can be preferentially reoriented under an applied electric field, particularly at low temperature when the ions are less mobile.[36]

In our previous work, we showed that lowering the temperature below $T= 200$ K can effectively suppress the ionic motion within hybrid perovskites devices.[9] By cooling methylammonium lead iodide ($CH_3NH_3PbI_3$) field effect transistors at temperatures below $T= 200$ K, wherein the effects of the ionic motion are strongly reduced due to lowered ionic mobility,[9, 30] we achieved a more balanced charge transport and higher effective mobility. Moreover, at low temperatures we could observe light emission from the channel of the transistor, realizing the first demonstration of perovskite light-emitting field-effect transistor (PeLEFET). Through observation of the electroluminescence (EL) signal we could also probe the movement of the recombination zone within the channel, which reveals gate-controlled charge injection and charge carrier transport mechanisms. However, low-temperature operation is unpractical for commercial applications of perovskite devices, which would mostly be required to operate under standard ambient conditions.

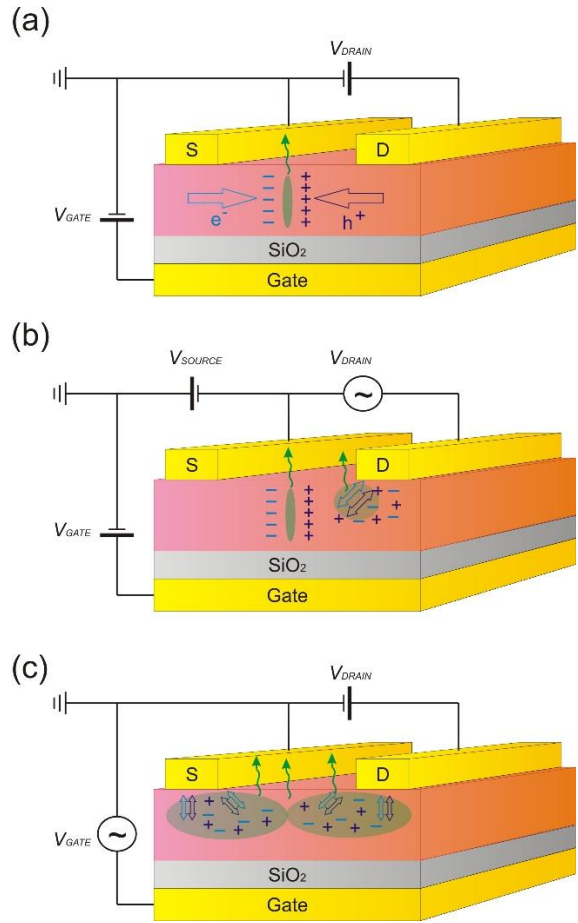

**Figure 1.** Schematics of perovskite light-emitting field effect transistors (PeLEFET) in different operation modes: (a) DC-Driven PeLEFET: holes and electrons are injected continuously from the source and the drain and recombine in a thin zone within the channel; (b) AC-driven drain PeLEFET: in addition to continuously injected charges due to the applied source bias, both electrons and holes are injected consecutively from the drain; (c) AC-driven gate PeLEFET: holes and electrons are injected consecutively from the source and the drain. Charge diffusion can be controlled by the lateral drain-source field.

In this work, we present a new device concept to overcome the detrimental effects of ionic drift and methylammonium cation polarization in perovskite light-emitting devices based on AC-modulation of the source-drain or gate bias in PeLEFETs. We show that the AC-driven PeLEFET is characterized by brighter and more uniform electroluminescence than DC-driven PeLEFETs at comparable applied voltages. Figure 1 shows different operational configurations of PeLEFET used in this work. In DC-driven PeLEFET[9] (Figure 1a), electrons and holes are steadily injected from source and drain and meet within the channel, giving rise to a thin recombination zone which can be moved by tuning the drain and gate bias. However, the DC-bias induced strong ionic migration and methylammonium cation ($MA^+$) polarization of the perovskite. In AC-driven drain PeLEFET (Figure 1b), the drain is driven by an AC voltage while a DC-bias applied to the gate and the source. Finally, in AC-driven gate PeLEFET (Figure 1c) an AC-bias is applied to the gate while a DC-bias is applied between the source and the drain. Application of AC-biases, especially at high frequencies, is expected to impede the ionic migration and the polarization of the methylammonium cation, if the driving frequency is faster than the response times of such phenomena, hence improving device efficiency. Indeed, we show that the AC-driven PeLEFET is characterized by brighter and more uniform electroluminescence than DC-driven PeLEFETs at comparable applied voltages. By tuning the drain bias and the amplitude of the gate bias we also observe uniform emission from the whole transistor channel. Importantly, high-frequency AC operation enables electro-luminescence emission at significantly higher temperatures, approaching room temperature.

## RESULTS AND DISCUSSION

We fabricated PeLEFETs in the top contact, bottom gate configuration, as shown in Fig. 1. The morphology of HOIPs is very sensitive to the deposition method, hence charge transport and electroluminescence characteristics are also closely tied to how the perovskite is deposited onto the substrate. In this work we used a modified version of "solvent engineering" method[55] to achieve a compact, smooth, crystalline and uniform $CH_3NH_3PbI_3$ film, obtaining good contact between the perovskite and the top contact source and drain electrodes. The resulting films are compact polycrystalline films with ~200 nm domains with a relatively smooth surface ($R_{RMS}$ = 6.79 nm), as seen from the AFM profile shown in Fig. S1b. $CH_3NH_3PbI_3$ crystallizes the characteristic room-temperature tetragonal phase with high purity and virtually absent unreacted $PbI_2$, as revealed by the X-ray diffraction analysis in Fig. S1b.

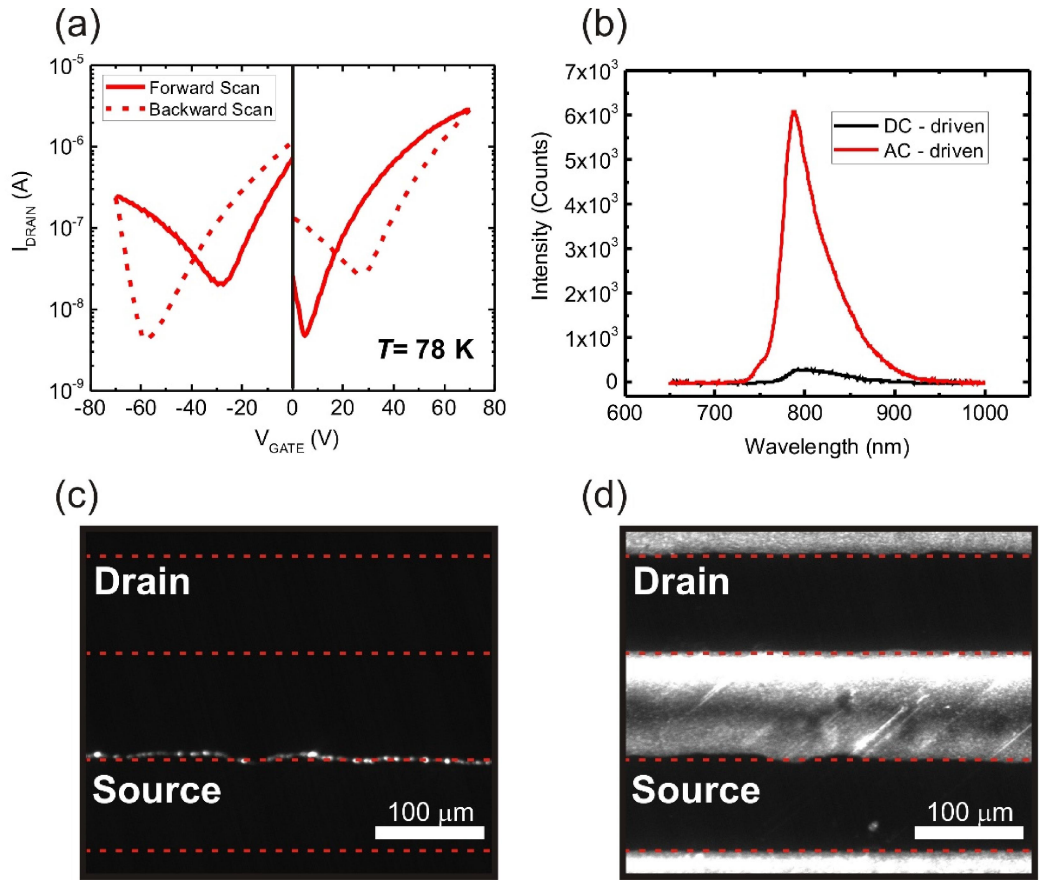

**Figure 2.** (a) Low-temperature transfer characteristics of the DC-driven PeLEFET. (b) Spectra of a PeLEFET driven in DC- and AC-mode at comparable voltages (DC: $V_{DRAIN}$=50 V, $V_{GATE}$=100 V; AC: $V_{DRAIN}$=50 V, $V_{GATE}$= ±100 V, 10 kHz square wave), showing the difference in electroluminescence between the two driving modes. Channel of the top contact bottom gate PeLEFET with electroluminescence at $T$=78 K from (c) the DC-driven PeLEFET ($V_{DRAIN}$=50 V, $V_{GATE}$=100 V), showing very low electroluminescence signal. (d) AC-driven PeLEFET with a 10 kHz square wave applied to the gate ($V_{DRAIN}$=50 V, $V_{GATE}$=±100 V).

Lowering the operating temperature below $T$=200 K decreases the ionic motion within the perovskite, significantly reducing its negative effects on device operation, hence recovering transistor-like behavior.[9, 30] Fig. 2a shows the transfer curves of the top contact PeLEFET at $T$=78 K. The PeLEFET has clear ambipolar behavior, which allows controlled injection of both charge carriers, hence tuning the position of the radiative recombination zone within the transistor channel,[9, 56-57] as shown in Fig. S2. Depending on the drain and gate biases applied, charge recombination and light emission occur either close to the drain or source electrode or within the transistor channel itself. Hysteresis in the current-voltage characteristics is still present even at low temperatures, and can be attributed not only to remnant ionic motion, but also to the presence of traps in the material.[30, 58-59] Thanks to the improved charge injection in the top contact configuration,[60] mobilities of 0.11 cm$^2$V$^{-1}$s$^{-1}$ for electrons and 0.025 cm$^2$V$^{-1}$s$^{-1}$ for holes were achieved at $T$=78 K, which are higher than those previously reported for CH$_3$NH$_3$PbI$_3$ devices with the same precursors and stoichiometry, yet significantly lower than theoretically predicted values.[5, 9, 47] For both electrons and holes, the mobility decreases slightly at higher temperatures, up until $T$=178 K, when it starts to decrease exponentially (Fig. S3). This is consistent with previous theoretical and experimental observations of the dependence of charge carrier mobility on the transition from orthogonal to tetragonal phase in methylammonium lead iodide.[7, 9, 30] when the perovskite is in the tetragonal phase, the polarization of the methylammonium cations causes energetic disorder at low temperatures (below ~250 K), hindering charge transport, while at higher temperatures the effect of ionic motion becomes the dominant factor affecting charge carrier mobility.[30]

Another approach to mitigate the negative effects of ionic drift and methylammonium cations polarization on charge transport and electroluminescence is the AC-driven PeLEFET. Figs. 2c-d show the comparison between the PeLEFET driven in DC-mode and AC-driven gate mode at comparable voltages. The AC-driven PeLEFET exhibits higher electroluminescence than DC-driven PeLEFET. When operated in DC-mode (Fig. 2c), at $V_{DRAIN}$=50 V and $V_{GATE}$=100 V, the PeLEFET electroluminescence is barely visible and radiative recombination occurs only in few spots, forming a narrow emission line close to the source electrode. On the other hand, when the PeLEFET gate is driven in AC-mode (Fig. 2d) with comparable voltages, $V_{DRAIN}$=50 V and $V_{GATE}$=±100 V by a 10 kHz square wave, the observed electroluminescence is strikingly higher, and light is emitted both from the edges of the drain and the source electrodes, and within the transistor channel. Fig. 2b shows the emission spectra of the PeLEFET device driven in DC- and AC-mode with voltages used in Figs. 2c-d. The maximum of the electroluminescence spectrum in AC-mode (at 10 kHz) is 20 times higher than in DC-mode. The increment in electroluminescence can be attributed to the reduction of ionic drift and MA$^+$ polarization, as well as to improved carrier injection. DC-bias causes the drift of iodine ion vacancies and respective ions, which eventually accumulate on one side of the device. This induces an electric field that screens the gating field, thus impeding charge transport and radiative recombination. On the other hand, if an AC-bias, such as a symmetric square wave, is applied to the gate, the movement of ion vacancies is restricted to an oscillation about the original point of origin, preventing accumulation of vacancies on one single side of the device. The same principle applies to lead and methylammonium ionic vacancies. Moreover, driving the bias at sufficiently high frequencies is also expected to avoid the effects of distortions induced by MA$^+$ polarization, if the period of the applied bias wave is faster than the MA$^+$ cation reorientation time. Moreover, in addition to reducing the effects of ionic motion and polarization effects, carrier injection is expected to improve due to *space-charge field-assisted* charge injection.[61] This effect might significantly contribute to the enhancement of the electroluminescence observed in the AC-driven devices. When the gate is driven with AC-bias, both charge carrier types are sequentially injected from the same electrode rather than one type of charge carrier being continuously injected (either from source or drain). During a half-period of the bias wave, one charge carrier type accumulates within the perovskite, inducing a space-charge field near the injecting electrode that aids injection of the other charge carrier type during the following half-period. A similar effect is also expected if the AC-bias is applied to the drain electrode instead of the gate.

The electroluminescence observed in the AC-driven PeLEFET (Figs. 2c-d), appears to originate not only within the channel of the transistor, but also from beneath the PeLEFET drain and source electrodes. This phenomenon has been previously observed in organic AC-driven LEFETs,[61-69] and was attributed to *AC field-induced electroluminescence* (AIFEL).[62-64] The operation of the field-effect transistor is fundamentally based on the principles of the metal-insulator-semiconductor (MIS) diode.[70] In our devices, both gate-source and gate-drain electrode pairs can be viewed as two separate MIS-diodes where one electrode is in contact with the semiconductor and acts as the injecting electrode (either the source or the drain electrode) while the other electrode, separated by an insulating layer, acts as the gate. When a bias is applied between the gate and the injecting electrode, holes or electrons are sequentially injected and accumulated within the semiconducting layer - not only within the transistor channel but also all around the electrodes. This phenomenon was also observed in purely capacitive electroluminescent devices with organic semiconductor active materials.[64, 71-73]

The spatial position of the carrier recombination (light emission) zone of the AC-driven PeLEFET can be controlled by modifying the gate or source drain voltages of the device, similarly to its DC-driven counterpart. This can be achieved in both cases of AC-driven drain or gate electrodes. As shown in Fig. 3a, by applying an alternating bias (±50 V, 10 kHz) to the drain electrode and keeping the gate bias constant (+20 V), the emission zone of the PeLEFET moves through the channel while sweeping the source-bias. By varying the source bias from +30 to +80 V, the recombination zone moves from the source (Fig. 3a, top panel) to the middle of the channel (Fig. 3a, middle panel), to the drain electrode (Fig. 3a, bottom panel). It is important to note that, because of the AC-bias applied to the drain electrode, there is *always* an emission contribution due to the AIFEL effect between the drain and the gate. The position of the recombination zone can also be tuned by applying a square wave to the gate and sweeping the drain bias. This is shown in Fig. S4, where $V_{GATE}$=± 50 V (square wave, 10 kHz) and the drain bias is swept from 20 to 80 V. Here the AIFEL effect clearly results in electroluminescence emission from the edges of the drain and source electrodes. In addition to having a spatially tunable recombination zone, tuning the gate and drain bias of the AC-driven gate PeLEFET allows achieving bright emission from the entire transistor channel (Fig. 3b). By applying a small (positive) bias to the drain electrode ($V_{SOURCE}$=+10 V) electroluminescence is observed mainly from the edges of the source and drain electrodes (Fig. 3b, top panel). By making the drain bias steadily more negative, the recombination zone moves within the transistor channel (Fig. 3b, middle panel) and, if the drain bias is sufficiently negative, light is emitted uniformly from the whole channel (Fig. 3b, bottom panel). Hence, by applying an AC-bias to the gate electrode, the source/drain and the gate electrodes can be capacitively coupled; the lateral field between the source and the drain electrodes induces charge carriers to drift and recombine throughout the channel, overall resulting in a wide and uniform emission zone across the top contacts.

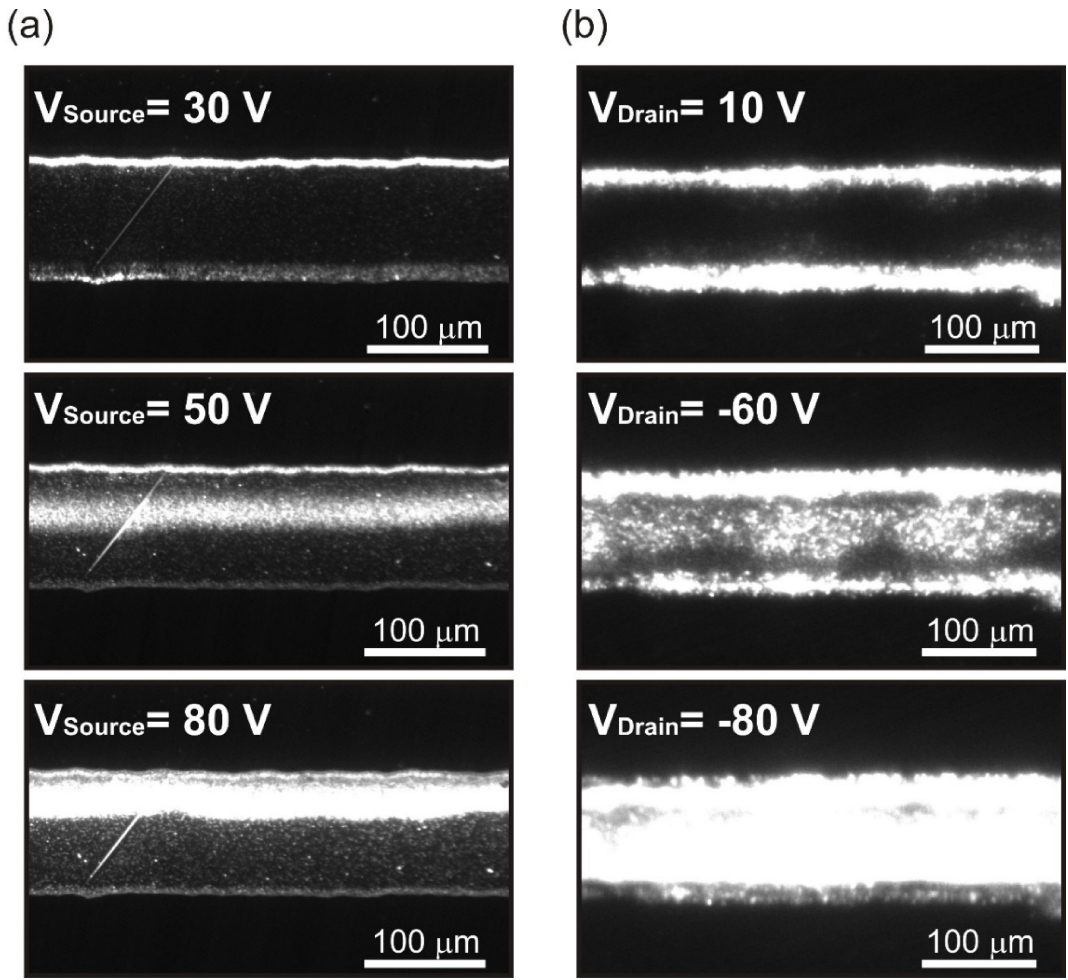

**Figure 3.** Low temperature ($T$=78 K) electroluminescence from the AC-driven PeLEFET in different operating conditions. (a) AC-driven drain: $V_{DRAIN}$=±50 V (10 kHz square wave), $V_{GATE}$=20 V and, from top to bottom, $V_{SOURCE}$=30 V, $V_{SOURCE}$=50 V and $V_{SOURCE}$=80 V, respectively. (b) AC-driven gate: $V_{GATE}$=±60 V (10 kHz square wave) and, from top to bottom, $V_{DRAIN}$=10 V, $V_{DRAIN}$=-60 V and $V_{DRAIN}$=-80 V, respectively.

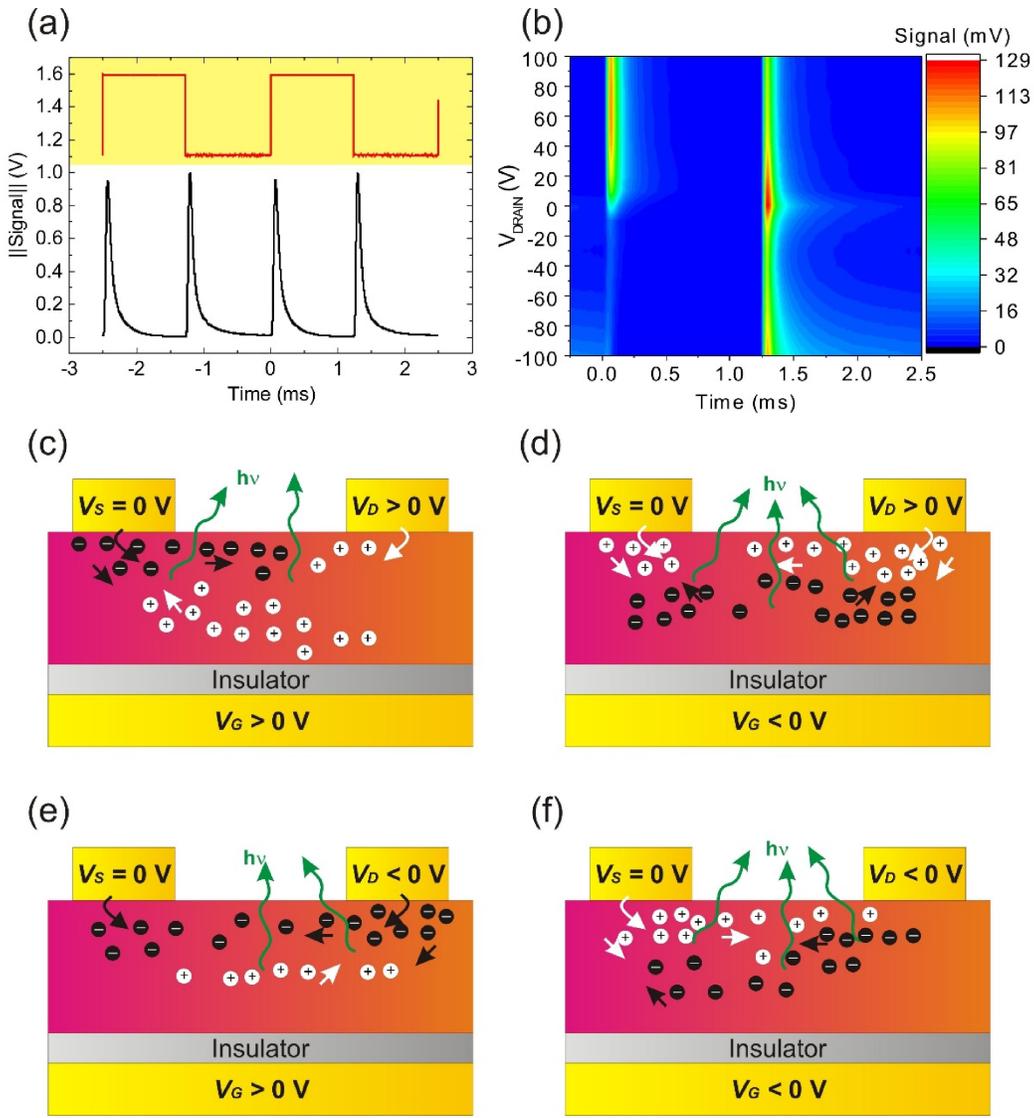

**Figure 4.** (a) Normalized plots of the applied square wave bias (top) and intensity of the electroluminescence signal response at 400 Hz, $V_{DRAIN}$=30 V, $V_{GATE}$=±50 V, and $T$=78 K (bottom). (b) Contour plot of the electroluminescence response of the AC-driven gate PeLEFET at 400 Hz, $V_{GATE}$=±50 V and $T$=78 K, for different drain bias. The polarity of the applied square wave was +50 V at $0 \leq t \leq 1.25$ ms and -50 V at $1.25 \leq t \leq 2.5$ ms. (c-f) Schematics of the operation of the AC-driven gate PeLEFET for positive drain bias and positive (c) or negative (d) gate bias, and for negative drain bias and positive (e) or negative (f) gate polarization.

The time evolution of the electroluminescence response of the PeLEFET to the square wave applied to the gate is shown in Figs. 4a-b. For sufficiently low frequencies, distinct electroluminescence peaks can be observed in response to positive and negative amplitudes of the square wave, as shown in Fig. 4a for a driving frequency of 400 Hz. The height of the electroluminescence peaks for positive and negative amplitudes varies according to the applied biases. By keeping the amplitude of the square wave constant, while varying the applied drain bias, the height and decay times of the peaks change significantly (Fig. 4b). For positive drain bias, electroluminescence occurs for both positive and negative polarity of the square wave applied to the gate, while for negative drain bias light is emitted only during the half-cycle with negative polarity. As the drain bias becomes more positive, the electroluminescence intensity in the half-cycle with positive polarity of the square wave increases, while in the half-cycle with negative polarity is slightly reduced. In the case of negative drain bias, the electroluminescence peak in the half-cycle with positive polarity is increasingly suppressed as the drain bias becomes more negative, while in the half-cycle with negative polarity it weakens up to $V_{DRAIN} \sim$-30 V, below which it becomes prominent and takes longer to decay, to the point where the decay time exceeds the period of the applied wave. This is consistent with uniform light emission observed throughout the channel at large negative drain bias amplitudes (Fig. 3b).

The electroluminescence peaks can be phenomenologically modelled using an asymmetric double sigmoidal function (Fig. S6) to describe the rise and decay times. The rise times of the peaks shows little dependence on applied drain bias, ranging between 9.6 and 11.3 µs when the square wave has positive amplitude, and between 8.4 and 9.5 µs when square wave has negative amplitude. The rise time of the signal is likely due to electroluminescence deriving from the AIFEL effect, where charges that are injected at the beginning of a half-cycle of the square wave could recombine with charge carriers of opposite sign accumulated in the preceding half-cycle. Here charge carriers need only to travel distances of ~100 nm between the source/drain electrodes and the gate. The transit time can then be approximated by $\tau \sim \frac{L}{\mu E} = \frac{L^2}{\mu V_{Eff}}$, where $L$ is the distance crossed, µ the charge carrier mobility and $V_{Eff}$ the effective voltage potential experienced by the charge carriers,[68] which is of the order of 15 ps for electrons and 67 ps for holes. The longer electroluminescence rise time observed in our experiments may arise from capacitive limitations, as the RC cut-off time of the drain/source-perovskite-oxide-gate MOS-capacitor is estimated to be around ~0.2 µs, corresponding to a cut-off frequency of 500 kHz, when the perovskite is in a low resistivity state.[74] The decay of the electroluminescence signal can be fitted with no less than to different decay times. The shorter decay time lies between ~35 and 45 µs, while the longer time lies between ~150 and 250 µs. The average ratio between the longer and shorter decay time is ~5, which is comparable to the ratio of measured hole and electron mobilities as $T$=78 K. This seems to indicate that the decay times are correlated to the mobility of charge carriers, with the shorter time relating to the transit time of electrons and the longer time relating to the transit time of holes. Since the decay times do not vary significantly with applied drain voltage, the effective voltage $V_{Eff}$ in the relation above appears to be almost independent of drain bias. This suggests that charge carriers involved in the recombination events experience screened gate and drain-source potentials.

From the above observations, we propose a physical description of the operation of the AC-driven gate PeLEFET for the different bias conditions, which is summarized in Figs. 4c-f. In all cases, light emission is expected from the edges of and under the electrodes due to the AIFEL-effect, given that charge carriers injected by the source and drain electrodes have high chance to meet and recombine with opposite charge carriers accumulated under or in the very close proximity of the injecting electrodes during the preceding half-cycle. With positive drain bias (Figs. 4c-d), electrons would tend to move towards the drain and holes towards the source. For positive AC gate bias polarity (Fig. 4c), electrons are injected from the source and drift towards both the gate and the drain, while holes are injected from the drain and drift mainly towards the source. Holes injected from the drain would also drift towards the source. Holes already present in the perovskite are driven towards the injecting electrodes, mainly the source, while electrons drift towards the gate dielectric, and accumulate at the interface. Hence, radiative recombination is observed primarily from the vicinity of the source electrode, but also in regions of the channel where electrons and holes meet. When the bias polarity is reversed (Fig. 4d), holes are injected from the source and the drain, accumulating close to the source and drifting mainly towards the gate dielectric. Holes already present in the perovskite are driven towards the gate dielectric and electrons towards the injecting electrodes, particularly towards the drain. Also in this case, light emission is expected to occur mainly near the electrodes, where electrons meet newly injected holes, but also from within the channel. This picture is consistent with the bright electroluminescence observed in the proximity of the electrodes for positive drain biases (except at very high fields), and with the existence of an additional recombination zone that can be moved within the channel, as shown in Fig. 3b (top panel) and Fig. S4. Moreover, this is also in agreement with time-resolved measurements, where large electroluminescence peak intensity was recorded for both positive and negative gate polarities. The relative peak amplitude at positive and negative gate polarity indicates that, at positive gate bias, electroluminescence increases for increasing positive drain bias (Fig. S5). This can be attributed to the fact that holes are the slowest charge carriers, hence take longer to drift back to the injecting electrodes once the gate polarity is reversed, hence resulting in higher concentration of holes than electrons at large positive bias. Overall, this increases the recombination yield when holes drift away from the gate-insulator interface, that is in the positive gate polarity half-cycle. With negative drain bias (Fig. 4e-f), electrons would tend to move towards the source and holes towards the drain electrodes. For positive AC gate bias polarity (Fig. 4e), electrons injected from the source would drift towards the gate dielectric, while electrons injected from the drain would drift towards the gate dielectric and the source. Radiative recombination could then occur with carriers injected from the previous half-cycle drifting towards the electrodes, especially the drain. The electroluminescence peak during the positive polarization of the gate is very weak and decreases for increasingly negative drain bias. This indicates that relatively few holes accumulate in this regime, consistent with their lower mobility hence longer transit time compared to the electrons. For negative polarity of the AC gate bias (Fig. 4f), holes are injected from the source and drift towards the drain, while electrons injected from the drain move towards the source. In this case, the device is electron dominated and light emission is determined by holes injected from the source and recombining with electrons already present in the perovskite. From Fig. 3b (middle and bottom panels), at negative drain bias recombination seems to occur both near the electrodes and within the channel, indicating that holes effectively drift within the channel and recombine with electrons throughout. Moreover, for negative gate bias, the electroluminescence peak is much more intense and is long lived, according to the hole transit time in the device.

The electroluminescence spectra of the PeLEFET driven at different frequencies, from 100 to 700 kHz, are shown in Fig. 5a. The corresponding frequency dependence of integrated electroluminescence signal is shown in Fig. 5b. At low frequencies, up to ~1 kHz, the electroluminescence intensity is linearly dependent on frequency, and the decay time of the electroluminescence pulses is smaller than half period of the applied square wave (Fig. 4a). In this regime, the integrated emission intensity is simply proportional to the number of pulses per unit time (modulation frequency). As the electroluminescence pulses get closely spaced, relative to the intrinsic decay time of the pulses, the integrated electroluminescence signal saturates, and eventually declines when the period of the AC-bias wave is faster than the pulse rise time (corresponding to frequencies of 10-20 kHz). Overall, the electroluminescence signal never saturates or declines between 4 and ~500 kHz modulation frequency. This follows a similar trend to the rise of the dielectric loss in response to frequency for perovskites at low temperature,[30] which is possibly caused by the polarization of the $MA^+$ cations within the perovskite.[30, 39, 53-54, 75] Hence, by increasing the modulation frequency, the polarization of $MA^+$ cations is expected to reduce and have less impact on transport and light emission characteristics of the device. This explains why the AC-driven PeLEFET is more efficient than the DC-driven counterpart, especially at applied frequencies above 10 kHz, as observed in Figures 2b-d. At modulation frequencies higher than ~500 kHz, the rise time of the electroluminescence pulse saturates (up to ~1.5 MHz) and successively declines (Fig. S8). Above 4.5 MHz no electroluminescence signal could be recorded. We attribute the saturation and decline of the electroluminescence signal to the cut-off frequency of the RC equivalent circuit, which is estimated to be around 500 kHz. In this high frequency regime, accumulation of charge carriers during each half-cycle reduces, resulting in saturation and eventual reduction of electroluminescence, up to the point where charge carriers cannot be effectively accumulated in the PeLEFET. Furthermore, the spectrum of the PeLEFET shows a slight dependence on modulation frequency, likely due to the modulation of self-absorption in methylammonium lead iodide for varying depth of the recombination zone.

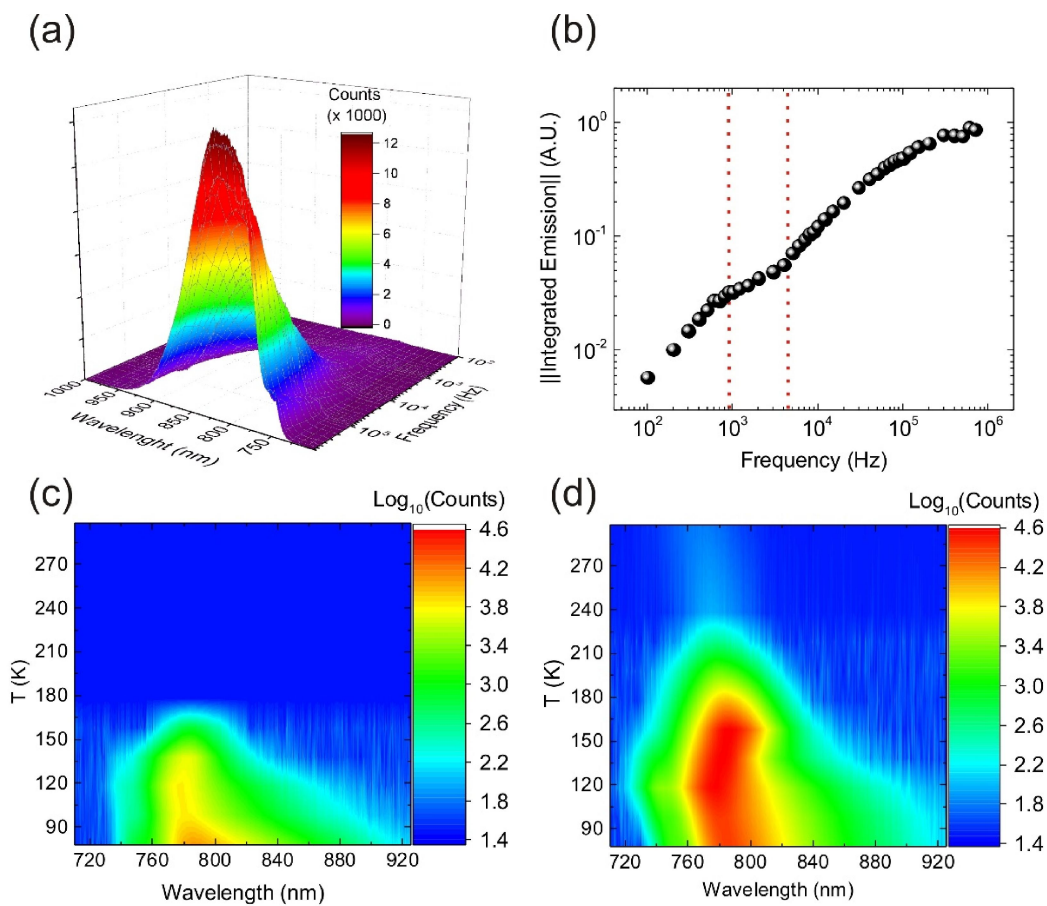

**Fig. 5.** (a) Low temperature ($T$=78 K) electroluminescence emission spectra of the PeLEFET with AC-driven gate (square wave) at different frequencies (from 100 to 700 kHz), $V_{DRAIN}$=-40 V and $V_{GATE}$=±60 V. (b) Normalized integrated emission of the electroluminescence peaks versus driving voltage frequency. The black dots are derived from the spectra at $V_{DRAIN}$=-40 V and $V_{GATE}$=±60 V ($T$=78 K). (c) Contour plot of the temperature dependent emission spectra of the DC-driven PeLEFET ($V_{DRAIN}$=150 V, $V_{GATE}$=100 V). (d) Contour plot of the temperature dependent emission spectra of the AC-driven PeLEFET ($V_{DRAIN}$=-80 V, $V_{GATE}$=±60 V, 100 kHz modulation frequency).

The reduction of ionic drift and MA$^+$ polarization in AC-driven PeLEFET significantly improves the electroluminescent characteristics of the device even at higher temperatures. As shown in Fig. 5c and previously reported,[9] the electroluminescence of DC-driven PeLEFETs becomes negligible above 180–200 K. The sudden decrease in electroluminescence efficiency at high temperatures is mainly attributed to the methylammonium cation polarization disorder, which becomes a significant source of energetic disorder in the tetragonal phase,[30] above ~160 K. This is corroborated by the fast decay of the electroluminescence peaks measured at low frequencies (400 Hz) above 160 K (Fig. S9), while at very high frequencies, such as 100 kHz, light emission is still significantly strong up to T~240 K. At high frequencies, a clear electroluminescence signal can be observed up to room temperature (Fig. 5d). The individual, normalized, spectra of the electroluminescence of the DC- and AC-driven PeLEFET at different temperatures are reported in Fig. S10. As it can be seen in Fig. S11, the temperature dependence of the integrated electroluminescence signal is similar to that of the fluorescence signal of $CH_3NH_3PbI_3$, especially at temperatures higher than 160 K. This suggests that emission from the AC-driven PeLEFET is closely related to the intrinsic properties of the pristine active material in the absence of electric field induced perturbations, such as ionic drift or MA$^+$ polarization. As expected, the total electroluminescence signal of the DC-driven PeLEFET seems to decay very quickly, similarly to the low frequency (400 Hz) electroluminescence pulses (Fig. S9), proving that high-frequency modulation of the PeLEFET effectively improves its performance at all temperatures, and enabling room temperature operation for the first time.

## CONCLUSIONS

In summary, we have demonstrated AC-driven, top-contact methylammonium lead iodide perovskite light-emitting field-effect transistors with controllable emission pattern and enhanced electroluminescence efficiency compared to DC-driven devices. The improved electroluminescence is attributed to minimization of ionic vacancy drift and methylammonium cation polarization within the perovskite active layer, as well as improved charge carrier injection due to space-charge field-assisted injection. By tuning the drain voltage and the amplitude of the gate bias we could achieve uniform emission from the entire PeLEFET channel. Moreover, high-frequency (100 kHz) AC operation enables electroluminescence emission at significantly higher temperatures than DC-driven PeLEFETs, approaching room temperature.

# EXPERIMENTAL DETAILS

**Perovskite synthesis and deposition.** Methylammonium lead iodide ($CH_3NH_3PbI_3$) thin films were prepared from 1 M precursor solutions of $CH_3NH_3I$ (Dyesol) and $PbI_2$ (99,99%, TCI) (molar ratio 1:1) in anhydrous DMF. Prior to the deposition of the perovskite the substrates (see FET fabrication) were cleaned with ultrasonication for 5 minutes in demineralized water, acetone and isopropanol and successively dried with compressed air. An oxygen plasma cleaning treatment was performed on the substrates, for 90 s, to improve the wetting of the surface and obtain flatter and homogeneous perovskite thin film. The DMF solution was spin-coated onto the substrates with a speed of 5,000 r.p.m. for 30 s. Toluene was drop-casted on the substrates 4 s after the start of the spin-coating program. The resulting film was finally annealed at 100 ºC for 15 min.

**Atomic force spectroscopy and X-ray diffraction.** The scanning probe microscope Cypher ES, Asylum Research, was used for atomic force microscopy measurements. The software Gwyddion was used for editing and plotting of the AFM images. The X-ray diffraction diffractograms were obtained using a diffractometer BRUKER D8 ADVANCE with Bragg-Brentano geometry employing Cu Ka radiation ($\lambda$= 1.54,056 Å), step increment of 0.02°, and 1 s of acquisition time.

**Field-effect transistor fabrication and characterization.** Heavily p-doped Si substrates were used as bottom gate electrode, with a 500 nm thick thermally grown silicon oxide layer as gate insulator with a capacitance of 6.9 $nFcm^{-2}$. The perovskite as deposited as described above and afterwards top contact field-effect transistors the source and drain electrodes (channel length: 100 μm, channel width: 1 mm) were deposited by thermal evaporation through a shadow mask on top of the previously deposited perovskite layer. The metal used as top contact were either gold or silver. The field-effect transistors were measured in vacuum ($10^{-3}$ mbar) and dark at different temperatures ranging from 78 K to 300 K using a liquid nitrogen-cooled Linkam Stage (HFS600E-PB4/PB2) with LNPT95 System Controller. The transistor DC-characteristics, with forward and reverse scans, were acquired with Agilent B2902A Precision Source/Measure Unit. The mobility of the electrons was determined using the conventional equations for metal-oxide–semiconductor transistors in saturated regime.[70] The AC-driven measurements were conducted applying a square wave bias on the PeLEFET gate electrode, using a Rigol DG5101 Function/Arbitrary Waveform Generator coupled with a Falco Systems WMA-300 High-Voltage amplifier.

**Electroluminescence characterization.** Optical images and recordings were acquired using a PCO.edge 3.1 sCMOS camera coupled to an optical microscope. Spectra were collected using an Avantes Avaspec ULS-RS-TEC. The time-dependent electroluminescent response of the PeLEFET with the applied bias was collected by using a Newport 818-UV photodiode connected to a Stanford Research Systems SR570 - Low noise current preamplifier and a LeCroy Wavesurfer 104MXs-B Oscilloscope. The signal trigger was obtained from the Rigol DG5101 Function/Arbitrary Waveform Generator.

# ASSOCIATED CONTENT

**Supporting Information**.

Figure S1. AFM and XRD measurements on methylammonium lead iodide.

Figure S2. Light emission in the DC-driven PeLEFET.

Figure S3. Temperature dependence of the charge carrier mobility of the DC-driven PeLEFET.

Figure S4. Electroluminescence of the PeLEFET with AC-driven gate.

Figure S5. EL-response of the PeLEFET with AC-driven gate at different drain biases.

Figure S6. Fitting of the time-resolved EL-response of the PeLEFET to an applied square wave bias.

Figure S7. Rise and decay times of the EL-peaks of the AC-driven gate PeLEFET.

Figure S8. Decay of the EL-signal at high frequencies

Figure S9. Contour plot of time-resolved EL-signal of the PeLEFET at different temperatures.

Figure S10. Temperature dependence of the electroluminescence spectra.

Figure S11. Plot of the normalized total electroluminescence for the AC- and DC-driven PeLEFET and the fluorescent signal of the perovskite from 78 to 298 K.

# AUTHOR INFORMATION

**Corresponding Author**

*Email: csoci@ntu.edu.sg

**Notes**

The authors declare no competing financial interest.

# ACKNOWLEDGEMENTS

Research was supported by the Singapore Ministry of Education (Grant Nos. MOE2016-T1-1-164 and MOE2011-T3-1-005) and the Singapore National Research Foundation (CRP Award No. NRF-CRP14-2014-03).

# Supplementary Information

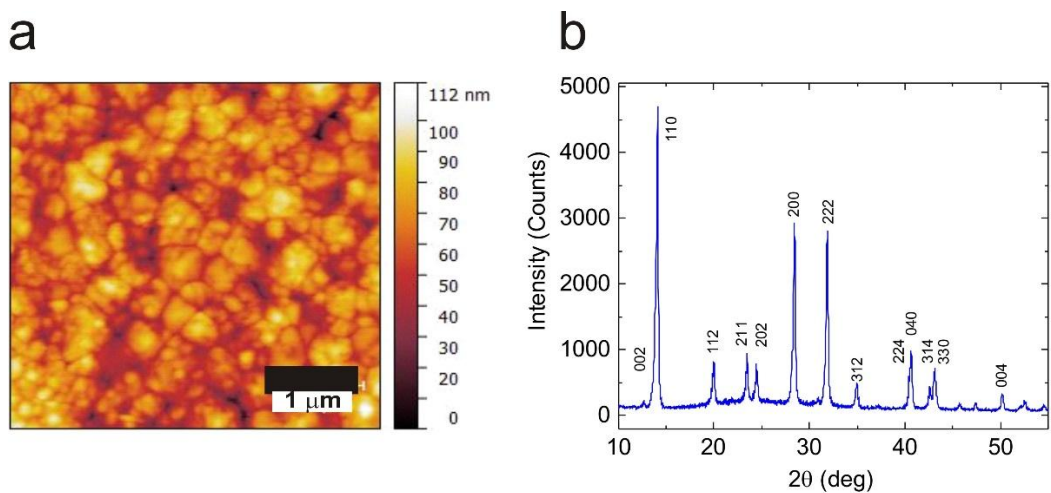

**Figure S1. AFM and XRD characterization of methylammonium lead iodide films. (**a) Atomic force microscopy height profile of the $CH_3NH_3PbI_3$ perovskite film. (b) X-ray diffraction pattern of $CH_3NH_3PbI_3$ film on $SiO_2/Si(p++)$ substrate, confirming the tetragonal structure of the perovskite and space group *I4/mcm*.

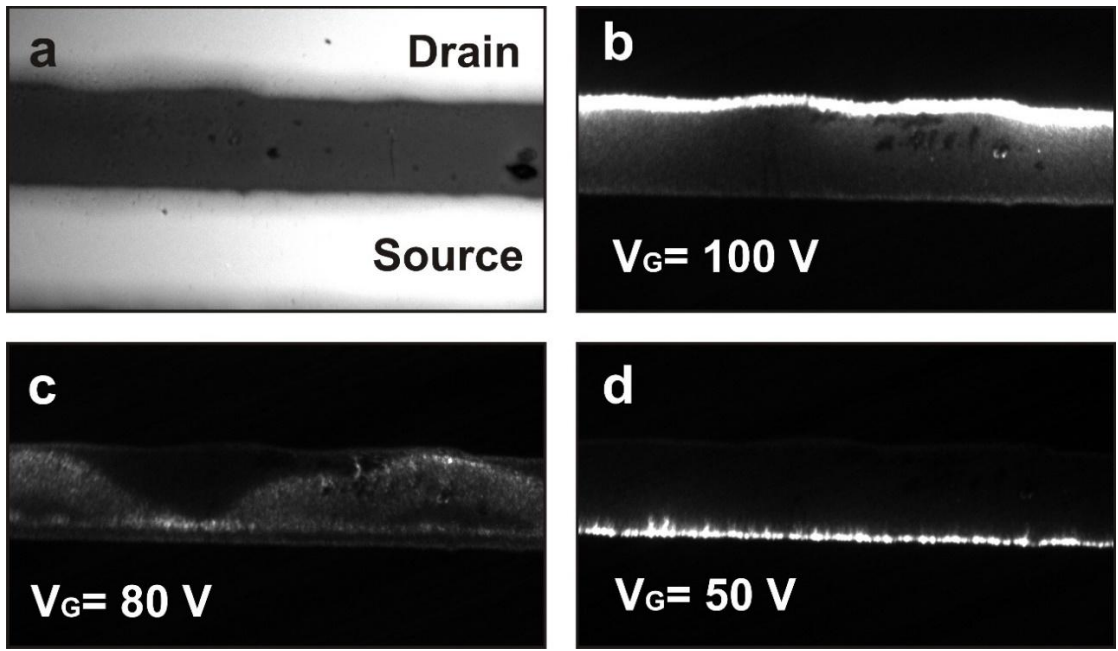

**Figure S2. Light emission from DC-driven PeLEFET.** a) Optical microscope image of the PeLEFET channel, with source and drain electrodes. a-d) Electroluminescence patterns recorded from the DC-driven PeLEFET with fixed $V_{DRAIN}$=150 V and different gate biases: b) $V_{GATE}$=100 V; c) $V_{GATE}$=80 V and d) $V_{GATE}$=50 V. The spatial position of the radiative recombination zone can be effectively moved within the PeLEFET channel by varying the gate voltage.

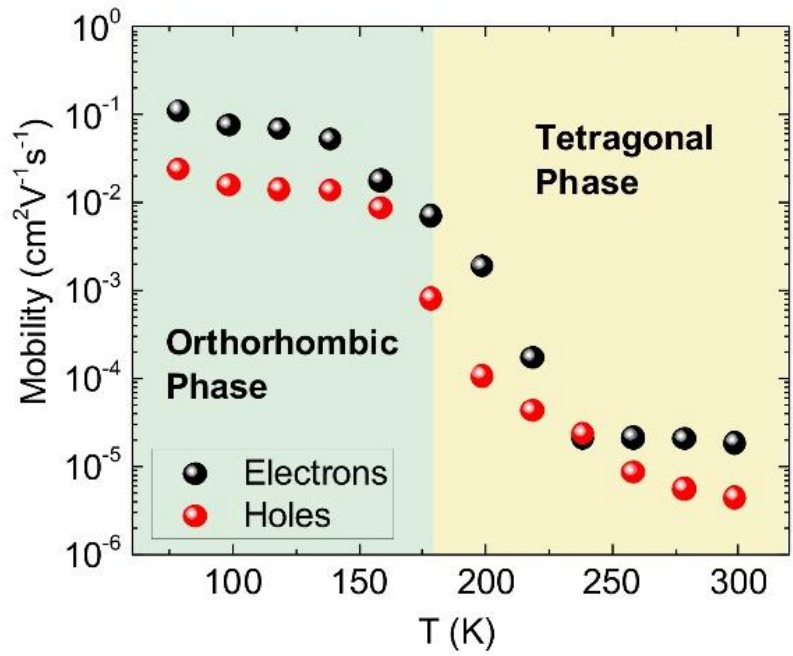

**Figure S3**. **Temperature dependence of the field-effect charge carrier mobility extracted from DC-driven PeLEFET.** Charge carrier mobility of methylammonium lead iodide PeLEFET ($L$= 100 µm, $W$= 1 mm) at $V_{DRAIN}$= 100 V, $V_{GATE}$= 60 V between 78 K and 298 K.

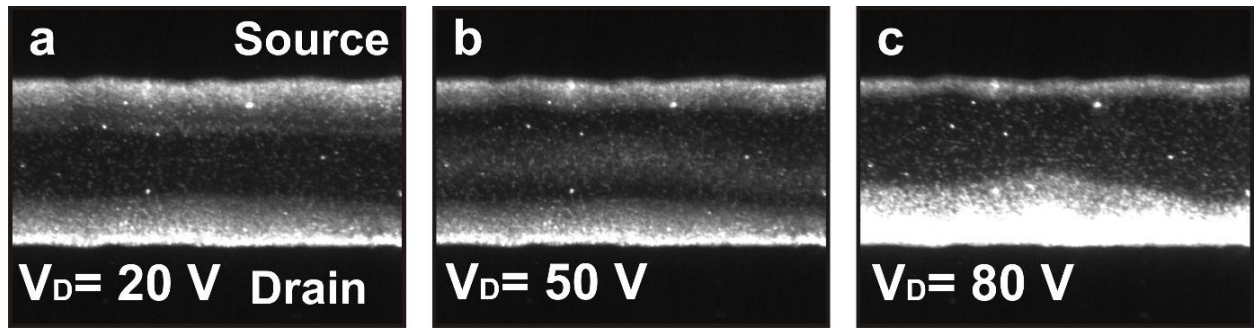

**Figure S4**. **Light emission from AC-driven PeLEFET.** Electroluminescence patterns recorded from the PeLEFET with AC-driven gate bias of $V_{GATE}$= ±50 V (square wave, 10 kHz), and different drain biases: a) $V_{DRAIN}$= 20 V; b) $V_{DRAIN}$=50 V and c) $V_{DRAIN}$= 80 V. By modulating the drain voltage, the position of the recombination zone can be shifted from the source to the drain electrodes. In addition to light emitted from within the transistor channel, the edges of source and drain electrodes always appear bright due to *AC field-induced electroluminescence* (AIFEL) induced by the capacitive coupling between the source/drain electrodes and the gate.

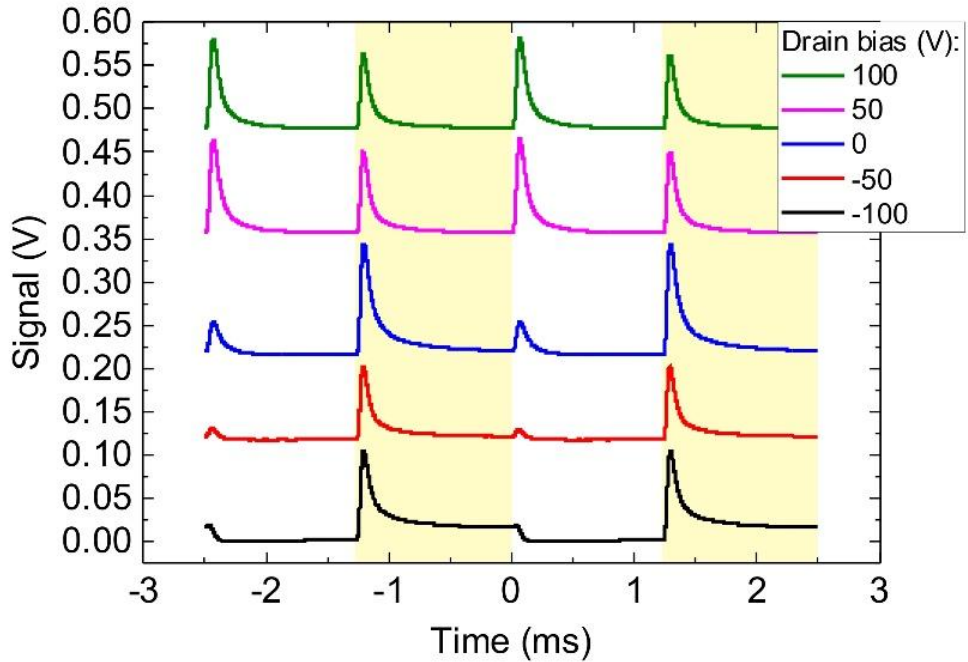

**Figure S5**. **Time-dependent electroluminescence response of the PeLEFET.** Time-resolved electroluminescence traces obtained with AC-driven gate bias of $V_{GATE}$=±50 V at 400 Hz (square wave) and different drain biases, from $V_{DRAIN}$=-100 V to $V_{DRAIN}$=100 V (see legend). Each cycle of the bias square wave produces two electroluminescence peaks, corresponding to the positive (white region) and negative (yellow region) polarity of the bias.

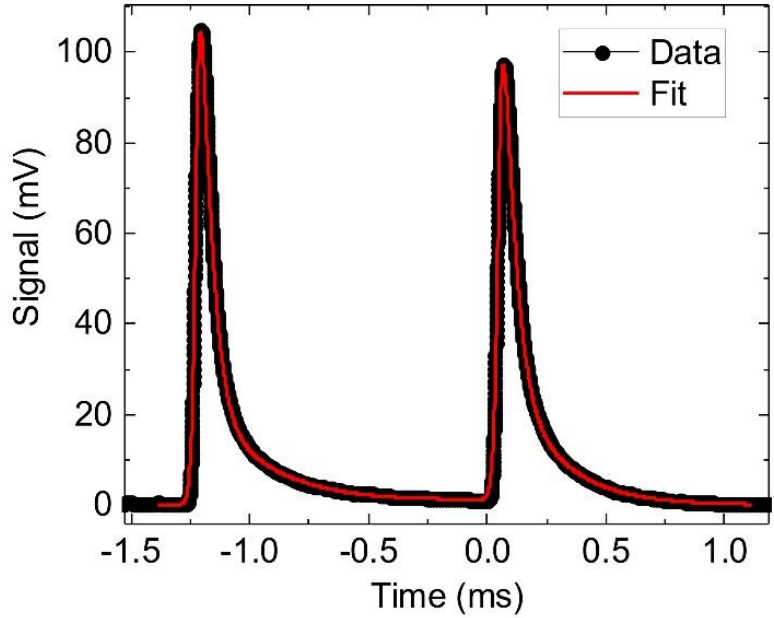

**Figure S6. Determination of the rise and decay times of the electroluminescence pulses upon application of a square wave bias.** Time resolved electroluminescence pulses obtained with AC-modulated gate bias of $V_{GATE}=\pm50$ V at a frequency of 400 Hz (square wave) and $V_{DRAIN}=30$ V. Fitting of the electroluminescence waveforms was carried out using the OriginPro2015 fitting application with a Levenberg–Marquardt algorithm. Fittings of the response in the negative polarity region (from -1.25 to 0 ms) and in the positive polarity region (from 0 to 1.25 ms) of the square wave were carried out independently. The intensity of the peaks *I(t)* in response to the applied square-wave bias was described phenomenologically using an asymmetric double sigmoidal function with one rising time and two decay times:

$$I(t) = \left( \frac{A}{1 + e^{-\frac{t-t_0}{\tau_R}}} + y_0 \right) \left[ 1 - \left( \left( \frac{B_1}{1 + e^{-\frac{t-t_0}{\tau_{D1}}}} \right) + \left( \frac{B_2}{1 + e^{-\frac{t-t_0}{\tau_{D2}}}} \right) + \left( \frac{B_3}{1 + e^{-\frac{t-t_0}{\tau_{D3}}}} \right) \right) \right]$$

Here *A* is a parameter that describes the peak amplitude of the rising function, $t_0$ describes the shift of the peak with respect of the origin, and $y_0$ represents the staring baseline of the sigmoidal function (which might be non-zero if the electroluminescence of the previous half-cycle has not fully decayed to zero). The parameters $t_R$, $t_{D1}$, $t_{D2}$ and $t_{D3}$ represent the rising time constant and the three decay time constants, respectively. The rising time is expected to correlate with injection of charge carriers into the devices, while the decay times with diffusion, hence mobility, of electrons and holes. Lastly the parameters $B_1$, $B_2$ and $B_3$ are the weight of the decay processes. While the parameters $A$, $y_0$, and $t_0$ were fixed based on notable points in the data sets, the amplitudes, rise and decay times of the electroluminescence waveforms were determined by the fitting optimization algorithm.

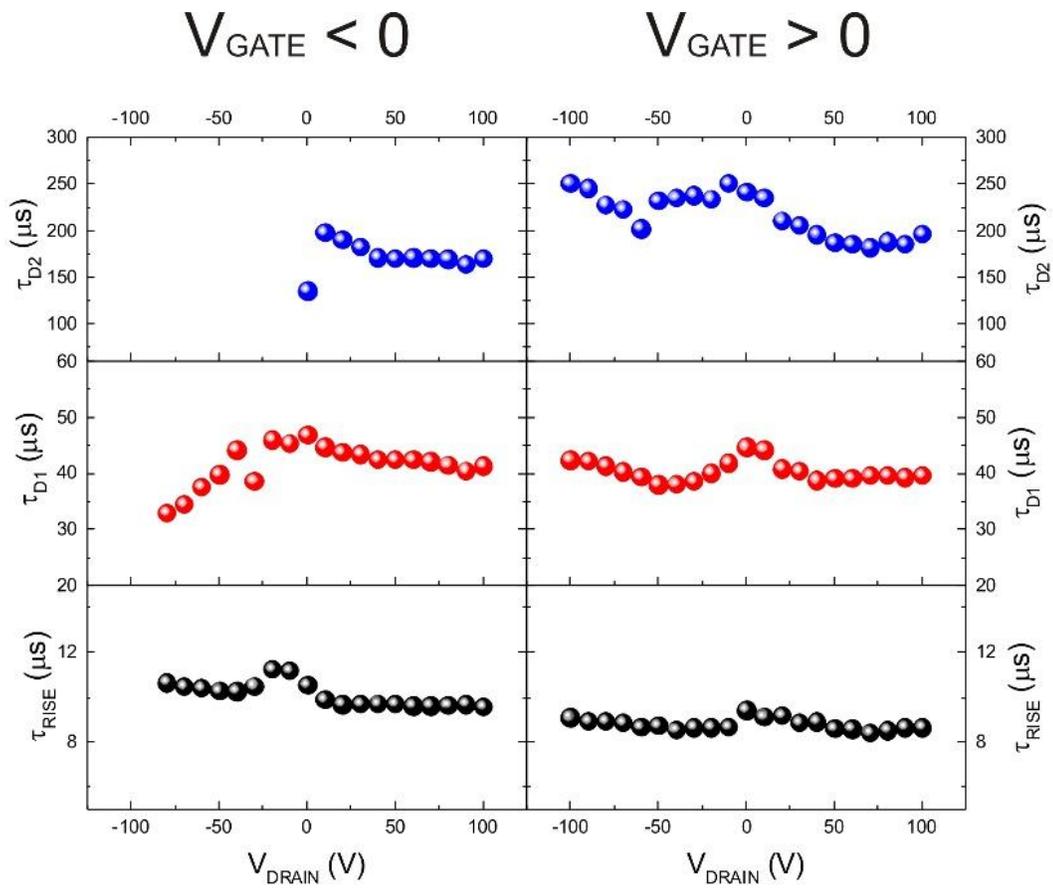

**Figure S7**. **Electroluminescence rise and decay times in PeLEFET with AC-driven gate.** The values, extracted using the model described in Figure S6, are given for the regimes of the applied square wave when the gate bias is negative ($V_{GATE}$=-50 V, left) and positive (of $V_{GATE}$=+50 V right).

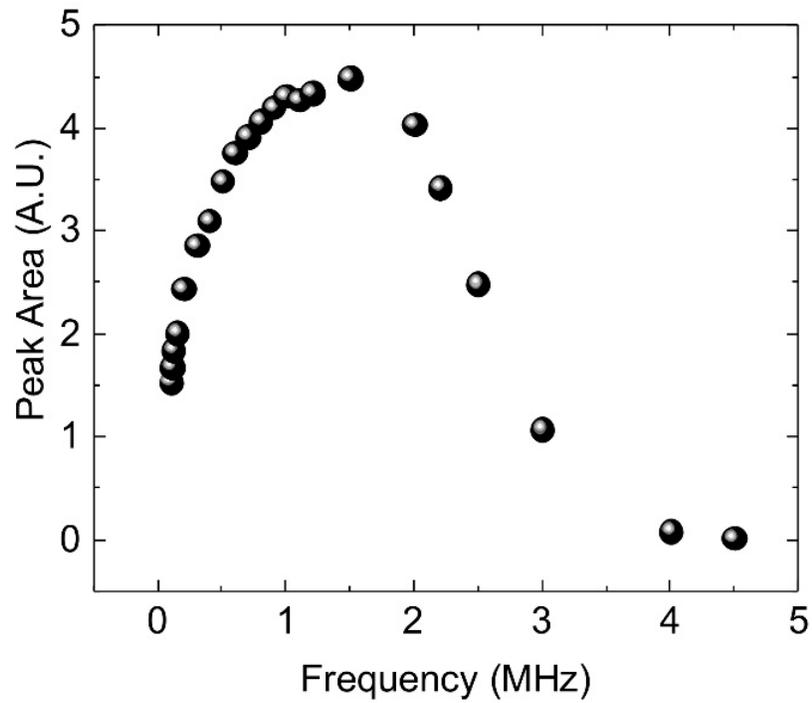

**Figure S8. Saturation and decay of the integrated electroluminescence signal at high modulation frequencies.** Integrated electroluminescence signal versus gate bias modulation frequency ($V_{GATE}$=± 60 V, square wave) at $V_{DRAIN}$ =-30 V. The electroluminescence signal saturates at modulation frequency of ~1.5 MHz, then quickly decays at higher frequencies.

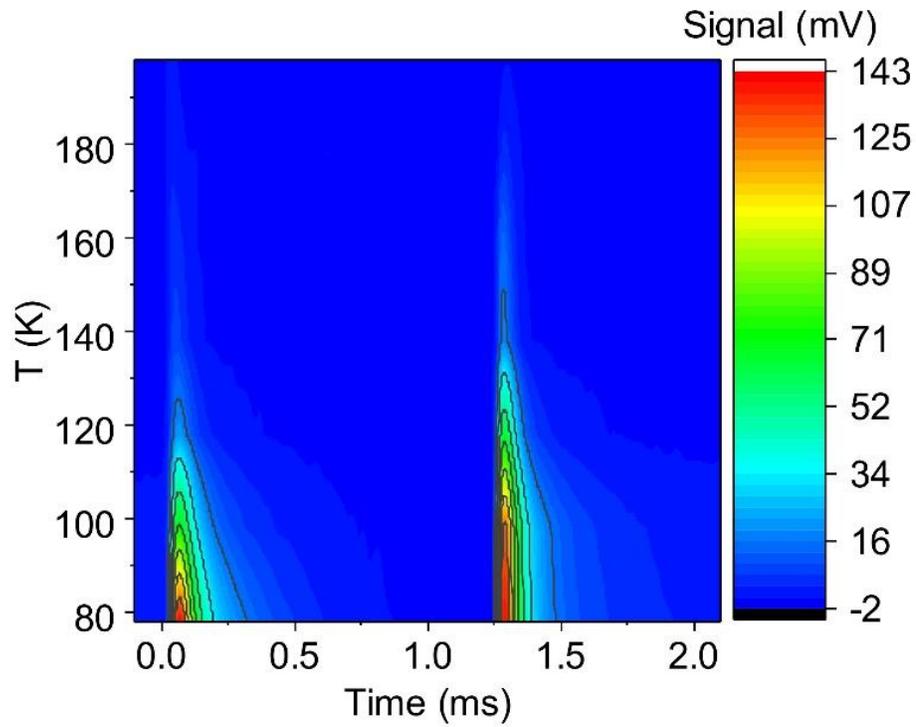

**Figure S9**. **Contour plot of the time-resolved electroluminescence traces at different temperatures.** Contour plot of the time-resolved response of the electroluminescence of the PeLEFET to the applied AC-bias ($V_{DRAIN}$= 40 V, $V_{GATE}$= ±50 V at 400 Hz square wave) at different temperatures.

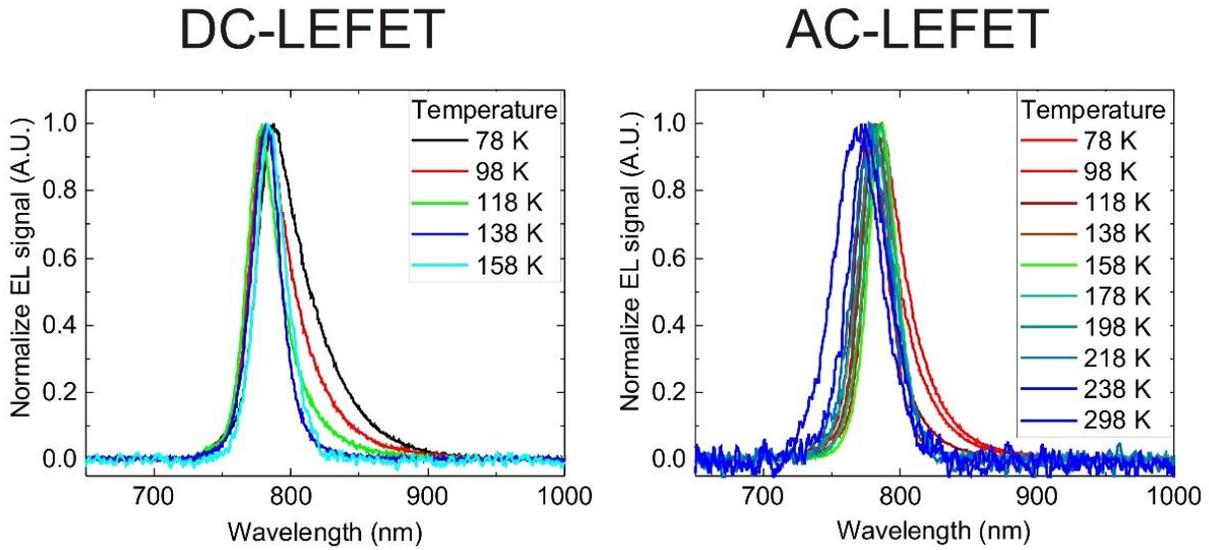

**Fig. S10. Temperature dependence of the electroluminescence spectra.** Normalized spectra of the total electroluminescence of the of methylammonium lead iodide PeLEFET in DC-mode (left panel, $V_{DRAIN}$= 150 V, $V_{GATE}$= 110 V) and in AC-mode (right panel, $V_{DRAIN}$= -80 V, $V_{GATE}$= ±60 V, 100 kHz square wave) at different temperatures. No significant signal was detected for the DC-LEFET above 158 K.

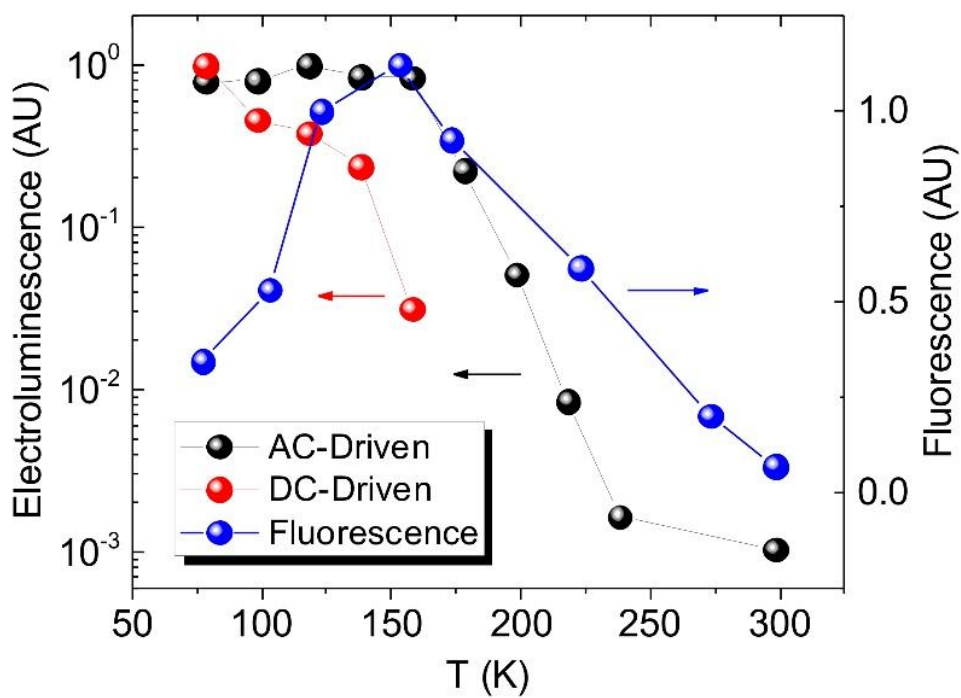

**Fig. S11**. **Temperature dependence of integrated electroluminescence and fluorescence signals.** Normalized total electroluminescence of the of methylammonium lead iodide PeLEFET in DC-mode ($V_{DRAIN}$= 150 V, $V_{GATE}$= 110 V) and in AC-mode ($V_{DRAIN}$= -80 V, $V_{GATE}$= ±60 V, 100 kHz square wave) and fluorescent signal from the methylammonium lead iodide film ($\lambda_{EXT}$= 400 nm).